# Exact Five-Loop Renormalization Group Functions of $\phi^4$-Theory with $O(N)$-Symmetric and Cubic Interactions. Critical Exponents up to $\epsilon^5$

H. Kleinert and V. Schulte-Frohlinde

Institut für Theoretische Physik

Freie Universität Berlin

Arnimallee 14 D - 14195 Berlin

**Abstract**

The renormalization group functions are calculated in $D = 4 - \epsilon$ dimensions for the $\phi^4$-theory with two coupling constants associated with an $O(N)$-symmetric and a cubic interaction. Divergences are removed by minimal subtraction. The critical exponents $\eta$, $\nu$, and $\omega$ are expanded up to order $\epsilon^5$ for the three nontrivial fixed points $O(N)$-symmetric, Ising, and cubic. The results suggest the stability of the cubic fixed point for $N \geq 3$, implying that the critical exponents seen in the magnetic transition of three-dimensional cubic crystals are of the cubic universality class. This is in contrast to earlier three-loop results which gave $N > 3$, and thus Heisenberg exponents. The numerical differences, however, are less than a percent making an experimental distinction of the universality classes very difficult.

**1** Many experimentally observable features of critical phenomena are described correctly by a scalar quantum field theory with a $\phi^4$-interaction. Field theoretic renormalization group techniques [1] in $D=4-\epsilon$ dimensions [2, 3, 4] combined with Borel resummation of the resulting $\epsilon$-expansions [5] have led to accurate determinations of the critical exponents of all $O(N)$ universality classes. An extension of the $O(N)$-symmetric theory by an interaction $\Sigma_i \phi_i^4$ makes the theory applicable to magnetic systems in which the $O(N)$ symmetry is broken and the magnetization is restricted to point along the edges or the diagonals of a hypercube in $N$ dimensions. The extended theory interpolates between an $O(N)$-symmetric and a cubic system. The $O(N)$-symmetric and the cubic fixed point interchange stability depending on $N$. For $N < N_c$, the first is stable, for $N > N_c$, the second is stable (see Fig. 1).

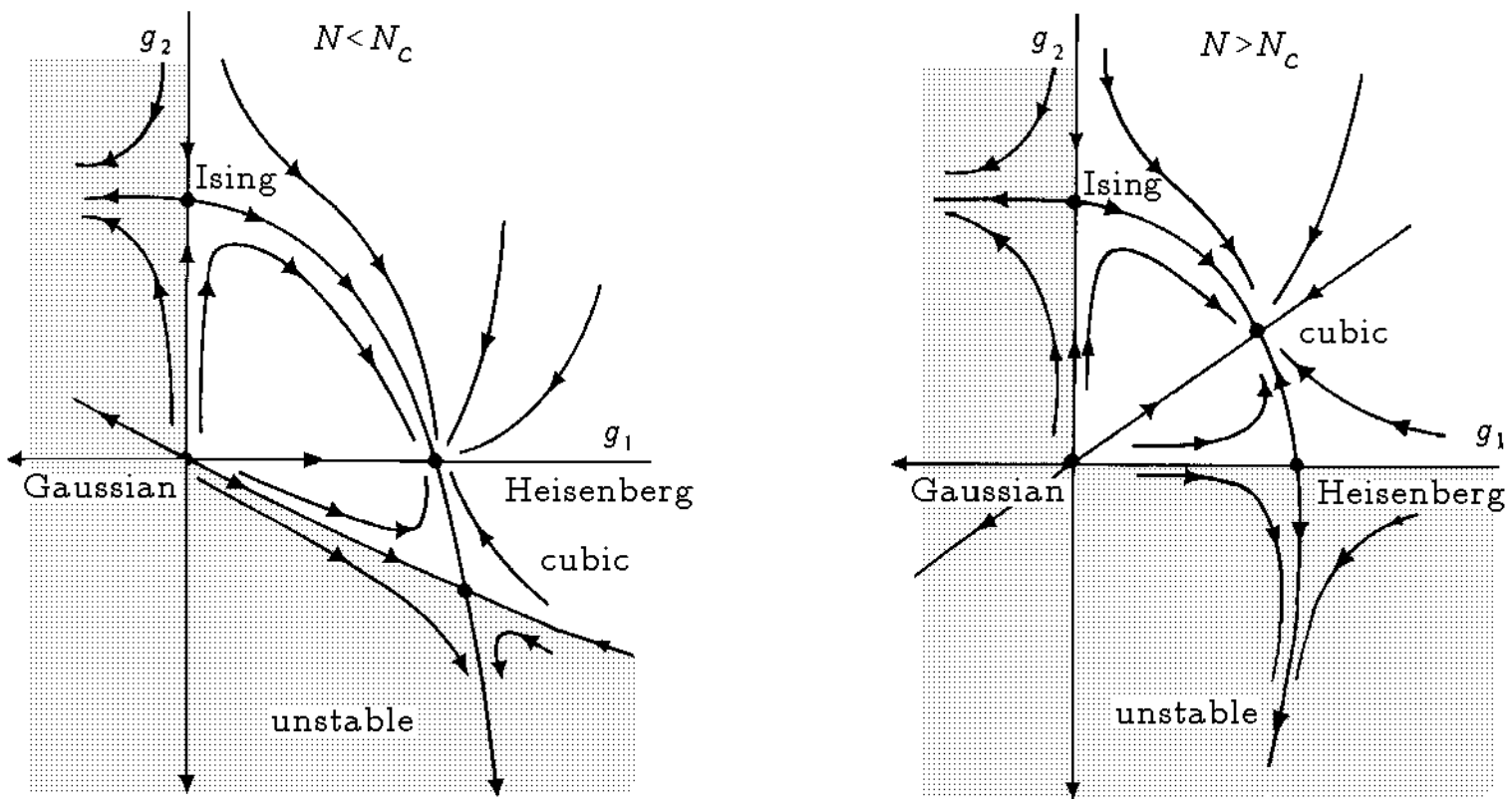

Figure 1: The Stability of the fixed points in the $\phi^4$-theory with $O(N)$-symmetric and cubic coupling for $N < N_c$ and $N > N_c$. Our results are compatible with $N_c = 3$.

From previous work based on three-loop calculations it is known that $N_c$ must lie somewhere between 3 and 4 [6, 7, 8]. Thus, for $N < N_c$, there is symmetry restoration: Although $O(N)$ symmetry is broken by the second interaction, it is restored by the fluctuations at the critical point. The physically most interesting value of $N$ is 3, where the $O(N)$-symmetric fixed point characterizes the critical behavior of the classical Heisenberg model of magnetism. The previous result $3 < N_c < 4$ implied that all magnetic systems with cubic symmetry occurring in nature show an $O(3)$-symmetric critical behavior.

The five-loop results to be presented in this letter suggest that the critical value $N_c$ lies *below* $N = 3$, so that the cubic fixed point may govern the critical behavior of magnetic transitions in cubic crystals. Due to the vicinity of the Heisenberg fixed point, however, the difference in the critical exponents will be hard to measure experimentally.

It is well known now to calculate analytically the Feynman graphs of the $\phi^4$-theory in $4-\epsilon$ dimensions using dimensional regularization [9] and the minimal substraction scheme [10] in the five-loop approximation [11]–[17]. The $\epsilon$-expansions of the renormalization group (RG) functions were given in [18] up to $\epsilon^5$. Semiclassical considerations have revealed the large-order behavior of the perturbation series [19] and of the $\epsilon$-expansions [20]. The two informations together yield the presently known critical exponents via several resummation techniques [5].

**2** In this paper we calculate the five-loop $\epsilon$-expansions for a $\phi^4$-theory with two coupling constants with the euclidean action

$$\mathcal{A} = \int d^D x \left[ \frac{1}{2} \partial_\mu \phi_i \partial_\mu \phi_i + \frac{1}{2} m_B^2 \phi_i \phi_i + \frac{16\pi^2}{4!} (g_{1_B}\, T^1_{ijkl} + g_{2_B}\, T^2_{ijkl})\; \phi_i \phi_j \phi_k \phi_l \right], \tag{1}$$

where $\phi_i,\ i = 1, 2, \ldots, N$ is a real $N$-component field in $D = 4 - \epsilon$ dimensions and $m_B,\ g_{1B},\ g_{2B}$ are the bare mass and coupling constants, respectively. The tensors in the two interaction terms have the following symmetrized form:

$$T^1_{ijkl} = \frac{1}{3}(\delta_{ij}\delta_{kl} + \delta_{ik}\delta_{jl} + \delta_{il}\delta_{kj}), \tag{2}$$

$$T^2_{ijkl} = \delta_{ijkl} \equiv \begin{cases} 1\ , & i = j = k = l, \\ 0\ , & \text{else.} \end{cases} \tag{3}$$

The action is positive definite within the *classical stability wedge* bounded by the lines $g_1 + g_2 > 0,\ g_1 > 0$ and $N g_1 + g_2 > 0,\ g_1 < 0$. This is changed drastically by the fluctuations (see Fig. 1).

The renormalization conditions are:

$$\Gamma^{(2)}_{ij}(p) \sim \Gamma^{(2)}(p)\ \delta_{ij}\ , \tag{4}$$

$$\Gamma^{(4)}_{ijkl}\Big|_{\text{S.P.}} \sim \Gamma^{(4)}_1\, T^1_{ijkl} + \Gamma^{(4)}_2\, T^2_{ijkl}, \tag{5}$$

where S.P. denotes the symmetry point of the four-point function at which the incoming momenta ${p^1}_\mu,\ {p^2}_\mu,\ {p^3}_\mu,\ {p^4}_\mu$ satisfy $p^\alpha p^\beta = (4\delta^{\alpha\beta} - 1)$. Condition (4) implies that the $N$ field components $\phi_i$ have only a single wave function renormalization constant. This is a consequence of the symmetry of the action under reflection $\phi_i \to -\phi_i$ and under permutations of the $N$ field indices $i$. The same symmetry guarantees the condition (5) to be fullfilled to all orders,

It should be noted that for $N = 2, 3$, the combinations of $T^1_{ijkl}$ and $T^2_{ijkl}$ exhaust all tensors of rank 4 for which the theory has only one length scale [21]. For $N \geq 4$, more tensors are admissible without introducing new length scales [22]; but we shall only consider the above two tensors, for simplicity.

There are four renormalization constants $Z_A$ $(A = \phi, 2, 4_1, 4_2)$ which relate the bare mass $m_B$ and two coupling constants $g_{iB}$ of Eq.(1) to the corresponding physical parameters by

$$m_B^2 = \frac{Z_2}{Z_\phi} m^2 = Z_{m^2} m^2; \qquad g_{i_B} = \mu^\epsilon \frac{Z_{4i}}{(Z_\phi)^2} g_i \qquad \text{for } i = 1, 2, \tag{6}$$

where $\mu$ is a mass parameter introduced for dimensional regularization. The RG-functions are introduced in the usual way:

$$\beta_i(g_1, g_2) = \mu \partial_\mu g_i |_{g_{1B}, g_{2B}, m_B, \epsilon} = \mu \partial_\mu g_i \Big|_B, \tag{7}$$

$$\gamma(g_1, g_2) = \mu \partial_\mu \log Z_\phi^{1/2} |_{g_{1B}, g_{2B}, m_B, \epsilon} = \mu \partial_\mu \log Z_\phi^{1/2} \Big|_B, \tag{8}$$

$$\gamma_m(g_1, g_2) = \mu \partial_\mu \log m |_{g_{1B}, g_{2B}, m_B, \epsilon} = \mu \partial_\mu \log m \Big|_B. \tag{9}$$

The calculation of the contribution of each graph to the RG-functions proceeds as described in [18] and references therein. For each graph, we evaluate the corresponding contractions of the tensors $T^1_{ijkl}$ and $T^2_{ijkl}$, decomposing the result again with respect to these tensors.

After some work, we obtain the following expressions for the RG-functions up to five loops:

$$\beta_1(g_1, g_2) = -\epsilon\, g_1 + {g_1}^2\, \tfrac{N+8}{3} + g_1\, g_2\, 2 + {g_1}^3 \left[-N - \tfrac{14}{3}\right] - {g_1}^2\, g_2\, \tfrac{22}{3} - g_1\, {g_2}^2\, \tfrac{5}{3} +$$

$$ {g_1}^4 \left[N^2\, \tfrac{11}{72} + N\left(\tfrac{461}{108} + \tfrac{20\,\zeta(3)}{9}\right) + \tfrac{370}{27} + \tfrac{88\,\zeta(3)}{9}\right] + {g_1}^3\, g_2 \left[\tfrac{79\,N}{36} + \tfrac{659}{18} + \tfrac{64\,\zeta(3)}{3}\right] + {g_1}^2\, {g_2}^2 \left[\tfrac{N}{24} + \tfrac{107}{4} + 8\,\zeta(3)\right] + g_1\, {g_2}^3\, 7 +$$

$$ {g_1}^5 \left[N^3 \tfrac{5}{3888} + N^2\left(-\tfrac{395}{243} - \tfrac{14\,\zeta(3)}{9} + \tfrac{10\,\zeta(4)}{27} - \tfrac{80\,\zeta(5)}{81}\right) + N\left(-\tfrac{10057}{486} - \tfrac{1528\,\zeta(3)}{81} + \tfrac{124\,\zeta(4)}{27} - \tfrac{2200\,\zeta(5)}{81}\right)\right.$$

$$\left. - \tfrac{24581}{486} - \tfrac{4664\,\zeta(3)}{81} + \tfrac{352\,\zeta(4)}{27} - \tfrac{2480\,\zeta(5)}{27}\right] +$$

$$ {g_1}^4\, g_2 \left[N^2\left(\tfrac{7}{81} - \tfrac{\zeta(3)}{9}\right) + N\left(-\tfrac{1319}{54} - \tfrac{184\,\zeta(3)}{9} + \tfrac{38\,\zeta(4)}{9} - \tfrac{400\,\zeta(5)}{27}\right) - \tfrac{15967}{81} - \tfrac{4856\,\zeta(3)}{27} + \tfrac{340\,\zeta(4)}{9} - \tfrac{2560\,\zeta(5)}{9}\right] +$$

$$ {g_1}^3\, {g_2}^2 \left[N\left(-\tfrac{301}{72} - \tfrac{35\,\zeta(3)}{9}\right) - \tfrac{13433}{54} - \tfrac{1456\,\zeta(3)}{9} + \tfrac{64\,\zeta(4)}{3} - \tfrac{2000\,\zeta(5)}{9}\right] +$$

$$ {g_1}^2\, {g_2}^3 \left[N\left(-\tfrac{25}{36} + \tfrac{\zeta(3)}{3}\right) - \tfrac{4867}{36} - 50\,\zeta(3) - 8\,\zeta(4) - \tfrac{160\,\zeta(5)}{3}\right] + g_1\, {g_2}^4 \left[-\tfrac{477}{16} - 3\,\zeta(3) - 6\,\zeta(4)\right] +$$

$$ {g_1}^6 \left[N^4\left(\tfrac{13}{62208} - \tfrac{\zeta(3)}{432}\right) + N^3 \left(\tfrac{6289}{31104} + \tfrac{26\,\zeta(3)}{81} - \tfrac{2\,\zeta^2(3)}{27} - \tfrac{7\,\zeta(4)}{24} + \tfrac{305\,\zeta(5)}{243} - \tfrac{25\,\zeta(6)}{81}\right) +\right.$$

$$ N^2 \left(\tfrac{50531}{3888} + \tfrac{8455\,\zeta(3)}{486} - \tfrac{59\,\zeta^2(3)}{81} - \tfrac{347\,\zeta(4)}{54} + \tfrac{7466\,\zeta(5)}{243} - \tfrac{1775\,\zeta(6)}{162} + \tfrac{686\,\zeta(7)}{27}\right) +$$

$$ N \left(\tfrac{103849}{972} + \tfrac{69035\,\zeta(3)}{486} + \tfrac{446\,\zeta^2(3)}{81} - \tfrac{2383\,\zeta(4)}{54} + \tfrac{66986\,\zeta(5)}{243} - \tfrac{7825\,\zeta(6)}{81} + 343\,\zeta(7)\right)$$

$$\left. + \tfrac{17158}{81} + \tfrac{27382\,\zeta(3)}{81} + \tfrac{1088\,\zeta^2(3)}{27} - \tfrac{880\,\zeta(4)}{9} + \tfrac{55028\,\zeta(5)}{81} - \tfrac{6200\,\zeta(6)}{27} + \tfrac{25774\,\zeta(7)}{27}\right] +$$

$$ {g_1}^5\, g_2 \left[N^3\left(\tfrac{161}{10368} - \tfrac{17\,\zeta(3)}{648} - \tfrac{\zeta(4)}{36}\right) + N^2 \left(\tfrac{59675}{15552} + \tfrac{170\,\zeta(3)}{27} - \tfrac{4\,\zeta^2(3)}{3} - \tfrac{19\,\zeta(4)}{4} + \tfrac{602\,\zeta(5)}{27} - \tfrac{50\,\zeta(6)}{9}\right) +\right.$$

$$N\left(\frac{5723}{27}+\frac{21560\,\zeta(3)}{81}-\frac{190\,\zeta^2(3)}{27}-\frac{2339\,\zeta(4)}{27}+\frac{4046\,\zeta(5)}{9}-\frac{4075\,\zeta(6)}{27}+\frac{1274\,\zeta(7)}{3}\right)$$

$$+\frac{537437}{486}+\frac{116759\,\zeta(3)}{81}+\frac{3148\,\zeta^2(3)}{27}-\frac{10177\,\zeta(4)}{27}+\frac{75236\,\zeta(5)}{27}-\frac{24050\,\zeta(6)}{27}+\frac{11564\,\zeta(7)}{3}\Bigg]+$$

$$g_1^4\,g_2^2\left[N^2\left(-\frac{1921}{10368}+\frac{763\,\zeta(3)}{648}-\frac{17\,\zeta(4)}{36}+\frac{5\,\zeta(5)}{9}\right)+N\left(\frac{270749}{2592}+\frac{9230\,\zeta(3)}{81}-\frac{232\,\zeta^2(3)}{27}-\frac{4841\,\zeta(4)}{108}+\frac{2045\,\zeta(5)}{9}-\frac{1450\,\zeta(6)}{27}+\frac{245\,\zeta(7)}{3}\right)\right.$$

$$\left.+\frac{1314497}{648}+\frac{171533\,\zeta(3)}{81}+\frac{1384\,\zeta^2(3)}{27}-\frac{23105\,\zeta(4)}{54}+\frac{96794\,\zeta(5)}{27}-\frac{25400\,\zeta(6)}{27}+\frac{14210\,\zeta(7)}{3}\right]+$$

$$g_1^3\,g_2^3\left[N\left(\frac{30277}{1296}+\frac{344\,\zeta(3)}{27}-\frac{25\,\zeta(4)}{6}+\frac{208\,\zeta(5)}{9}\right)+\frac{2281727}{1296}+\frac{37789\,\zeta(3)}{27}-\frac{544\,\zeta^2(3)}{9}-\frac{337\,\zeta(4)}{3}+\frac{17444\,\zeta(5)}{9}-\frac{1600\,\zeta(6)}{9}+2352\,\zeta(7)\right]+$$

$$g_1^2\,g_2^4\left[N\left(\frac{26171}{6912}-\frac{77\,\zeta(3)}{48}+\frac{7\,\zeta(4)}{8}-\frac{4\,\zeta(5)}{3}\right)+\frac{1336801}{1728}+\frac{5495\,\zeta(3)}{12}-\frac{190\,\zeta^2(3)}{3}+\frac{141\,\zeta(4)}{2}+\frac{1145\,\zeta(5)}{3}+\frac{575\,\zeta(6)}{3}+441\,\zeta(7)\right]+$$

$$g_1\,g_2^5\left[\frac{158849}{1152}+\frac{1519\,\zeta(3)}{24}-18\,\zeta^2(3)+\frac{65\,\zeta(4)}{2}+2\,\zeta(5)+75\,\zeta(6)\right]\;, \tag{10}$$

$$\beta_2(g_1,g_2)=-\epsilon\,g_2+3\,g_2^2+4\,g_1\,g_2-g_2^3\,\frac{17}{3}-g_1\,g_2^2\,\frac{46}{3}-g_1^2\,g_2\,\frac{5\,N+82}{9}+g_2^4\left[\frac{145}{8}+12\,\zeta(3)\right]+$$

$$g_1\,g_2^3\left[\frac{131}{2}+48\,\zeta(3)\right]+g_1^2\,g_2^2\left[\frac{17\,N}{24}+\frac{325}{4}+64\,\zeta(3)\right]+g_1^3\,g_2\left[-N^2\,\frac{13}{108}+N\left(\frac{92}{27}+\frac{16\,\zeta(3)}{9}\right)+\frac{821}{27}+\frac{224\,\zeta(3)}{9}\right]+$$

$$g_2^5\left[-\frac{3499}{48}-78\,\zeta(3)+18\,\zeta(4)-120\,\zeta(5)\right]+g_1\,g_2^4\left[-\frac{1004}{3}-387\,\zeta(3)+96\,\zeta(4)-600\,\zeta(5)\right]+$$

$$g_1^2\,g_2^3\left[N\left(-\frac{19}{24}-\frac{19\,\zeta(3)}{3}+4\,\zeta(4)\right)-\frac{10661}{18}-724\,\zeta(3)+184\,\zeta(4)-\frac{3440\,\zeta(5)}{3}\right]+$$

$$g_1^3\,g_2^2\left[N^2\left(\frac{1}{6}+\frac{\zeta(3)}{9}\right)+N\left(-\frac{508}{27}-\frac{218\,\zeta(3)}{9}+12\,\zeta(4)-\frac{160\,\zeta(5)}{9}\right)-\frac{12349}{27}-\frac{5312\,\zeta(3)}{9}+\frac{440\,\zeta(4)}{3}-960\,\zeta(5)\right]+$$

$$g_1^4\,g_2\left[N^3\left(-\frac{29}{1296}+\frac{\zeta(3)}{27}\right)+N^2\left(-\frac{7}{162}-\frac{8\,\zeta(3)}{9}+\frac{4\,\zeta(4)}{9}\right)+\right.$$

$$\left.N\left(-\frac{3479}{162}-\frac{560\,\zeta(3)}{27}+\frac{68\,\zeta(4)}{9}-\frac{280\,\zeta(5)}{9}\right)-\frac{19679}{162}-168\,\zeta(3)+40\,\zeta(4)-\frac{7280\,\zeta(5)}{27}\right]+$$

$$g_2^6\left[\frac{764621}{2304}+\frac{7965\,\zeta(3)}{16}+45\,\zeta^2(3)-\frac{1189\,\zeta(4)}{8}+987\,\zeta(5)-\frac{675\,\zeta(6)}{2}+1323\,\zeta(7)\right]+$$

$$g_1\,g_2^5\left[\frac{1067507}{576}+\frac{35083\,\zeta(3)}{12}+288\,\zeta^2(3)-\frac{3697\,\zeta(4)}{4}+5920\,\zeta(5)-2100\,\zeta(6)+7938\,\zeta(7)\right]+$$

$$g_1^2\,g_2^4\left[N\left(-\frac{16223}{3456}+\frac{2947\,\zeta(3)}{72}-17\,\zeta^2(3)-\frac{151\,\zeta(4)}{4}+\frac{290\,\zeta(5)}{3}-\frac{125\,\zeta(6)}{2}\right)\right.$$

$$\left.+\frac{3633377}{864}+\frac{125459\,\zeta(3)}{18}+\frac{2266\,\zeta^2(3)}{3}-2263\,\zeta(4)+14328\,\zeta(5)-\frac{15575\,\zeta(6)}{3}+19404\,\zeta(7)\right]+$$

$$g_1^3\,g_2^3\left[N^2\left(\frac{8213}{15552}-\frac{35\,\zeta(3)}{108}-\frac{2\,\zeta(4)}{3}+\frac{14\,\zeta(5)}{9}\right)+N\left(\frac{496159}{7776}+\frac{1309\,\zeta(3)}{6}-\frac{452\,\zeta^2(3)}{9}-\frac{4076\,\zeta(4)}{27}+478\,\zeta(5)-\frac{2450\,\zeta(6)}{9}+196\,\zeta(7)\right)\right.$$

$$\left.+\frac{9309907}{1944}+\frac{224804\,\zeta(3)}{27}+\frac{3032\,\zeta^2(3)}{3}-\frac{73018\,\zeta(4)}{27}+\frac{155692\,\zeta(5)}{9}-6300\,\zeta(6)+23912\,\zeta(7)\right]+$$

$$g_1^4\,g_2^2\left[N^3\left(\frac{127}{20736}-\frac{91\,\zeta(3)}{1296}+\frac{\zeta(4)}{18}\right)+N^2\left(-\frac{43295}{31104}+\zeta(3)-\frac{4\,\zeta^2(3)}{3}-\frac{121\,\zeta(4)}{24}+\frac{364\,\zeta(5)}{27}-\frac{50\,\zeta(6)}{9}\right)+\right.$$

$$N\left(\frac{11495}{54}+\frac{31598\,\zeta(3)}{81}-\frac{1045\,\zeta^2(3)}{27}-\frac{10729\,\zeta(4)}{54}+\frac{20917\,\zeta(5)}{27}-\frac{21425\,\zeta(6)}{54}+\frac{1960\,\zeta(7)}{3}\right)$$

$$\left.+\frac{1279979}{486}+\frac{784621\,\zeta(3)}{162}+\frac{18154\,\zeta^2(3)}{27}-\frac{83837\,\zeta(4)}{54}+\frac{275510\,\zeta(5)}{27}-\frac{98975\,\zeta(6)}{27}+\frac{43120\,\zeta(7)}{3}\right]+$$

$$g_1^5\,g_2\left[N^4\left(-\frac{61}{15552}-\frac{5\,\zeta(3)}{972}+\frac{\zeta(4)}{108}\right)+N^3\left(-\frac{3557}{46656}-\frac{151\,\zeta(3)}{972}-\frac{4\,\zeta(4)}{27}+\frac{8\,\zeta(5)}{81}\right)+\right.$$

$$N^2\left(\frac{111217}{23328}+\frac{2785\,\zeta(3)}{243}-\frac{92\,\zeta^2(3)}{81}-\frac{1055\,\zeta(4)}{162}+\frac{530\,\zeta(5)}{27}-\frac{950\,\zeta(6)}{81}+\frac{98\,\zeta(7)}{9}\right)+$$

$$N\left(\frac{95588}{729}+\frac{15742\,\zeta(3)}{81}-\frac{92\,\zeta^2(3)}{81}-\frac{6592\,\zeta(4)}{81}+\frac{34460\,\zeta(5)}{81}-\frac{14750\,\zeta(6)}{81}+490\,\zeta(7)\right)$$

$$\left.+\frac{389095}{729}+\frac{259358\,\zeta(3)}{243}+\frac{13288\,\zeta^2(3)}{81}-\frac{27166\,\zeta(4)}{81}+\frac{179696\,\zeta(5)}{81}-\frac{63500\,\zeta(6)}{81}+\frac{28420\,\zeta(7)}{9}\right]\;, \tag{11}$$

$$\gamma_2(g_1,g_2)=g_1^2\,\frac{N+2}{36}+g_1\,g_2\,\frac{1}{6}+g_2^2\,\frac{1}{12}-g_1^3\left[\frac{N^2}{432}+\frac{5\,N}{216}+\frac{1}{27}\right]-g_1^2\,g_2\left[\frac{N}{48}+\frac{1}{6}\right]-g_1\,g_2^2\,\frac{3}{16}-g_2^3\,\frac{1}{16}+$$

$$g_1^4\left[-\frac{5N^3}{5184}+\frac{5N^2}{324}+\frac{85N}{648}+\frac{125}{648}\right]+g_1^3g_2\left[-\frac{5N^2}{432}+\frac{5N}{24}+\frac{125}{108}\right]+g_1^2g_2^2\left[\frac{5N}{288}+\frac{145}{72}\right]+g_1g_2^3\,\frac{65}{48}+g_2^4\,\frac{65}{192}+$$
$$g_1^5\left[N^4\left(-\frac{13}{62208}+\frac{\zeta(3)}{3888}\right)+N^3\left(-\frac{187}{93312}-\frac{\zeta(3)}{972}\right)+N^2\left(-\frac{1459}{11664}+\frac{13\,\zeta(3)}{972}-\frac{5\,\zeta(4)}{162}\right)+\right.$$
$$\left.N\left(-\frac{1915}{2916}+\frac{13\,\zeta(3)}{162}-\frac{16\,\zeta(4)}{81}\right)-\frac{602}{729}+\frac{23\,\zeta(3)}{243}-\frac{22\,\zeta(4)}{81}\right]+$$
$$g_1^4g_2\left[N^3\left(-\frac{65}{20736}+\frac{5\,\zeta(3)}{1296}\right)+N^2\left(-\frac{185}{7776}-\frac{5\,\zeta(3)}{216}\right)+N\left(-\frac{395}{216}+\frac{20\,\zeta(3)}{81}-\frac{25\,\zeta(4)}{54}\right)-\frac{1505}{243}+\frac{115\,\zeta(3)}{162}-\frac{55\,\zeta(4)}{27}\right]+$$
$$g_1^3g_2^2\left[\frac{325N^2}{31104}+N\left(-\frac{4453}{3888}+\frac{23\,\zeta(3)}{216}-\frac{5\,\zeta(4)}{27}\right)-\frac{58177}{3888}+\frac{191\,\zeta(3)}{108}-\frac{130\,\zeta(4)}{27}\right]+$$
$$g_1^2g_2^3\left[N\left(-\frac{671}{3456}+\frac{\zeta(3)}{72}\right)-\frac{13741}{864}+\frac{67\,\zeta(3)}{36}-5\,\zeta(4)\right]+g_1g_2^4\left[-\frac{18545}{2304}+\frac{15\,\zeta(3)}{16}-\frac{5\,\zeta(4)}{2}\right]+g_2^5\left[-\frac{3709}{2304}+\frac{3\,\zeta(3)}{16}-\frac{\zeta(4)}{2}\right]\,, \qquad (12)$$

$$\gamma_m(g_1,g_2)=g_1\,\frac{2+N}{6}+g_2\,\frac{1}{2}-g_1^2\,\frac{5N+10}{36}-g_1g_2\,\frac{5}{6}-g_2^2\,\frac{5}{12}+$$
$$g_1^3\left[\frac{5N^2}{72}+\frac{47N}{72}+\frac{37}{36}\right]+g_1^2g_2\,\frac{37+5N}{8}+g_1g_2^2\,\frac{251+N}{48}+g_2^3\,\frac{7}{4}+$$
$$g_1^4\left[N^3\left(\frac{1}{15552}-\frac{\zeta(3)}{108}\right)+N^2\left(-\frac{947}{1944}-\frac{4\,\zeta(3)}{81}-\frac{5\,\zeta(4)}{54}\right)+N\left(-\frac{5777}{1944}-\frac{22\,\zeta(3)}{81}-\frac{16\,\zeta(4)}{27}\right)-\frac{7765}{1944}-\frac{34\,\zeta(3)}{81}-\frac{22\,\zeta(4)}{27}\right]+$$
$$g_1^3g_2\left[N^2\left(\frac{1}{1296}-\frac{\zeta(3)}{9}\right)+N\left(-\frac{421}{72}-\frac{10\,\zeta(3)}{27}-\frac{10\,\zeta(4)}{9}\right)-\frac{7765}{324}-\frac{68\,\zeta(3)}{27}-\frac{44\,\zeta(4)}{9}\right]+$$
$$g_1^2g_2^2\left[N\left(-\frac{1841}{864}-\frac{\zeta(3)}{6}-\frac{\zeta(4)}{3}\right)-\frac{9199}{216}-\frac{13\,\zeta(3)}{3}-\frac{26\,\zeta(4)}{3}\right]+$$
$$g_1g_2^3\left[N\left(-\frac{25}{72}+\frac{\zeta(3)}{6}\right)-\frac{4243}{144}-\frac{19\,\zeta(3)}{6}-6\,\zeta(4)\right]+g_2^4\left[-\frac{477}{64}-\frac{3\,\zeta(3)}{4}-\frac{3\,\zeta(4)}{2}\right]+$$
$$g_1^5\left[N^4\left(\frac{7}{124416}+\frac{17\,\zeta(3)}{7776}-\frac{\zeta(4)}{432}\right)+N^3\left(\frac{2831}{23328}+\frac{487\,\zeta(3)}{3888}-\frac{\zeta^2(3)}{243}+\frac{23\,\zeta(4)}{1296}-\frac{5\,\zeta(5)}{486}+\frac{25\,\zeta(6)}{486}\right)+\right.$$
$$N^2\left(\frac{291907}{93312}+\frac{1261\,\zeta(3)}{972}-\frac{149\,\zeta^2(3)}{486}+\frac{437\,\zeta(4)}{648}+\frac{2\,\zeta(5)}{243}+\frac{1475\,\zeta(6)}{972}\right)+$$
$$N\left(\frac{83137}{5832}+\frac{2011\,\zeta(3)}{324}-\frac{436\,\zeta^2(3)}{243}+\frac{1075\,\zeta(4)}{324}+\frac{50\,\zeta(5)}{243}+\frac{1850\,\zeta(6)}{243}\right)$$
$$\left.+\frac{49477}{2916}+\frac{1327\,\zeta(3)}{162}-\frac{194\,\zeta^2(3)}{81}+\frac{667\,\zeta(4)}{162}+\frac{8\,\zeta(5)}{27}+\frac{775\,\zeta(6)}{81}\right]+$$
$$g_1^4g_2\left[N^3\left(\frac{35}{41472}+\frac{85\,\zeta(3)}{2592}-\frac{5\,\zeta(4)}{144}\right)+N^2\left(\frac{113135}{62208}+\frac{1175\,\zeta(3)}{648}-\frac{5\,\zeta^2(3)}{81}+\frac{145\,\zeta(4)}{432}-\frac{25\,\zeta(5)}{162}+\frac{125\,\zeta(6)}{162}\right)+\right.$$
$$N\left(\frac{4675}{108}+\frac{95\,\zeta(3)}{6}-\frac{725\,\zeta^2(3)}{162}+\frac{85\,\zeta(4)}{9}+\frac{35\,\zeta(5)}{81}+\frac{6875\,\zeta(6)}{324}\right)$$
$$\left.+\frac{247385}{1944}+\frac{6635\,\zeta(3)}{108}-\frac{485\,\zeta^2(3)}{27}+\frac{3335\,\zeta(4)}{108}+\frac{20\,\zeta(5)}{9}+\frac{3875\,\zeta(6)}{54}\right]+$$
$$g_1^3g_2^2\left[N^2\left(-\frac{2045}{20736}+\frac{785\,\zeta(3)}{1296}-\frac{7\,\zeta(4)}{72}-\frac{7\,\zeta(5)}{54}\right)+N\left(\frac{362281}{10368}+\frac{3743\,\zeta(3)}{324}-\frac{61\,\zeta^2(3)}{27}+\frac{1493\,\zeta(4)}{216}-\frac{11\,\zeta(5)}{18}+\frac{775\,\zeta(6)}{54}\right)\right.$$
$$\left.+\frac{267737}{864}+\frac{47327\,\zeta(3)}{324}-\frac{1154\,\zeta^2(3)}{27}+\frac{8039\,\zeta(4)}{108}+\frac{155\,\zeta(5)}{27}+\frac{4675\,\zeta(6)}{27}\right]+$$
$$g_1^2g_2^3\left[N\left(\frac{26173}{2304}+\frac{71\,\zeta(3)}{48}-\frac{2\,\zeta^2(3)}{9}+\frac{25\,\zeta(4)}{12}-\frac{14\,\zeta(5)}{9}+\frac{25\,\zeta(6)}{9}\right)\right.$$
$$\left.+\frac{32003}{96}+\frac{627\,\zeta(3)}{4}-\frac{403\,\zeta^2(3)}{9}+\frac{475\,\zeta(4)}{6}+\frac{59\,\zeta(5)}{9}+\frac{3325\,\zeta(6)}{18}\right]+$$
$$g_1g_2^4\left[N\left(\frac{26171}{13824}-\frac{77\,\zeta(3)}{96}+\frac{7\,\zeta(4)}{16}-\frac{2\,\zeta(5)}{3}\right)+\frac{589141}{3456}+\frac{959\,\zeta(3)}{12}-\frac{45\,\zeta^2(3)}{2}+\frac{643\,\zeta(4)}{16}+\frac{19\,\zeta(5)}{6}+\frac{375\,\zeta(6)}{4}\right]+$$
$$g_2^5\left[\frac{158849}{4608}+\frac{1519\,\zeta(3)}{96}-\frac{9\,\zeta^2(3)}{2}+\frac{65\,\zeta(4)}{8}+\frac{\zeta(5)}{2}+\frac{75\,\zeta(6)}{4}\right]\,. \qquad (13)$$

**3** The three critical exponents $\eta$, $\nu$ and $\omega$ completely specify the critical behavior of the system [4]. They are determined from the RG-functions of the $\phi^4$-theory by going to the infrared-stable fixed points $(g_1^\star, g_2^\star)$. These are found from the zeros of the $\beta$-functions in the form offree $\epsilon$-expansions:

$$g_i^\star = g_i^\star(\epsilon) = \sum_{k=1}^{\infty} g_i^{(k)} \epsilon^k \ . \tag{14}$$

Besides the Gaussian fixed point ($g_1^\star = g_2^\star = 0$), there are the following three nontrivial fixed points, Ising ($g_1^\star = 0, g_2^\star = g_2^I$), symmetric or Heisenberg ($g_1^\star = g_1^H, g_2^\star = 0$) and cubic ($g_1^\star = g_1^C, g_2^\star = g_2^C$). For $N = 1$, the Heisenberg fixed point coincides with the Ising fixed point ($g_1^H = g_2^I$), whereas cubic ($g_1^C = -g_2^C$) and the Gaussian fixed point will have degenerated critical exponents. For $N = 2$, we will find degenerated exponents for the cubic and the Ising fixed point ($-g_2^C = g_2^I$). As $N$ increases, the cubic fixed point approaches the Heisenberg fixed point from below, crossing it at $N = N_c$, where $g_2^C$ changes its sign. For $N \to \infty$, the cubic fixed point moves into the Ising fixed point.

Since the $\epsilon$-expansions for the Ising and Heisenberg fixed points can be deduced from the series in Ref. [18], we exhibit the expansions only for the cubic fixed point

$$\begin{aligned}
g_1^C \ = \ & \epsilon\,\frac{1}{N} + \epsilon^2\,\frac{\left(-106+125\,N-19\,N^2\right)}{27\,N^3} + \\
& \epsilon^3 \left[\frac{22472}{729\,N^5} - \frac{45080}{729\,N^4} + \frac{38329}{972\,N^3} - \frac{41971}{5832\,N^2} - \frac{1955}{5832\,N} + \left(\frac{56}{9\,N^4} - \frac{28}{9\,N^3} - \frac{8}{3\,N^2} + \frac{8}{9\,N}\right)\zeta(3)\right] + \\
& \epsilon^4 \left[-\frac{5955080}{19683\,N^7} + \frac{5623300}{6561\,N^6} - \frac{5934115}{6561\,N^5} + \frac{8315992}{19683\,N^4} - \frac{3955061}{52488\,N^3} + \frac{113779}{104976\,N^2} - \frac{2987}{314928\,N}\right. \\
& \quad + \left(-\frac{29680}{243\,N^6} + \frac{46696}{243\,N^5} - \frac{12764}{243\,N^4} - \frac{7934}{243\,N^3} + \frac{2744}{243\,N^2} + \frac{110}{243\,N}\right)\zeta(3) \\
& \quad \left. + \left(\frac{32}{9\,N^4} - \frac{16}{9\,N^3} - \frac{16}{9\,N^2} + \frac{2}{3\,N}\right)\zeta(4) + \left(-\frac{80}{3\,N^5} + \frac{200}{27\,N^4} + \frac{80}{27\,N^3} + \frac{80}{9\,N^2} - \frac{80}{27\,N}\right)\zeta(5)\right] + \\
& \epsilon^5 \left[\frac{1767467744}{531441\,N^9} - \frac{6468340480}{531441\,N^8} + \frac{9496212881}{531441\,N^7} - \frac{14088835643}{1062882\,N^6} + \frac{43137004355}{8503056\,N^5}\right. \\
& \quad - \frac{7400332843}{8503056\,N^4} + \frac{2080479877}{68024448\,N^3} + \frac{337198481}{136048896\,N^2} - \frac{5795035}{136048896\,N} \\
& \quad + \left(\frac{4404512}{2187\,N^8} - \frac{3678896}{729\,N^7} + \frac{9044242}{2187\,N^6} - \frac{1239931}{1458\,N^5} - \frac{1694161}{4374\,N^4} + \frac{653341}{4374\,N^3} - \frac{182483}{34992\,N^2} + \frac{13883}{34992\,N}\right)\zeta(3) \\
& \quad + \left(-\frac{18020}{243\,N^6} + \frac{56137}{486\,N^5} - \frac{2213}{81\,N^4} - \frac{47369}{1944\,N^3} + \frac{3929}{486\,N^2} + \frac{217}{648\,N}\right)\zeta(4) \\
& \quad + \left(\frac{16960}{27\,N^7} - \frac{579260}{729\,N^6} + \frac{107902}{729\,N^5} - \frac{508}{27\,N^4} + \frac{84818}{729\,N^3} - \frac{26296}{729\,N^2} - \frac{340}{243\,N}\right)\zeta(5) \\
& \quad + \left(-\frac{2225}{81\,N^5} + \frac{1525}{162\,N^4} + \frac{125}{81\,N^3} + \frac{850}{81\,N^2} - \frac{100}{27\,N}\right)\zeta(6) + \left(\frac{-1078}{9\,N^6} + \frac{1225}{3\,N^5} - \frac{2450}{9\,N^4} + \frac{539}{9\,N^3} - \frac{343}{9\,N^2} + \frac{98}{9\,N}\right)\zeta(7) \\
& \quad \left. + \left(\frac{3136}{27\,N^7} - \frac{112}{N^6} - \frac{2834}{81\,N^5} + \frac{3089}{81\,N^4} + \frac{626}{81\,N^3} - \frac{296}{81\,N^2} - \frac{16}{27\,N}\right)\zeta^2(3)\right] + O[\epsilon^6] \ ,
\end{aligned} \tag{15}$$

$$\begin{aligned}
g_2^C \ = \ & \epsilon\,\frac{(N-4)}{3\,N} + \epsilon^2\,\frac{\left(424-534\,N+93\,N^2+17\,N^3\right)}{81\,N^3} + \\
& \epsilon^3 \left[-\frac{89888}{2187\,N^5} + \frac{187528}{2187\,N^4} - \frac{123707}{2187\,N^3} + \frac{90281}{8748\,N^2} + \frac{11713}{17496\,N} + \frac{709}{17496} + \left(-\frac{224}{27\,N^4} + \frac{16}{3\,N^3} + \frac{80}{27\,N^2} - \frac{32}{27\,N} - \frac{4}{27}\right)\zeta(3)\right] + \\
& \epsilon^4 \left[\frac{23820320}{59049\,N^7} - \frac{69389720}{59049\,N^6} + \frac{25018256}{19683\,N^5} - \frac{35478331}{59049\,N^4} + \frac{11944655}{118098\,N^3} + \frac{406721}{157464\,N^2} - \frac{511435}{944784\,N} + \frac{10909}{944784}\right. \\
& \quad + \left(\frac{118720}{729\,N^6} - \frac{200768}{729\,N^5} + \frac{23752}{243\,N^4} + \frac{2704}{81\,N^3} - \frac{10450}{729\,N^2} - \frac{56}{81\,N} - \frac{106}{729}\right)\zeta(3) \\
& \quad \left. + \left(-\frac{136}{27\,N^4} + \frac{28}{9\,N^3} + \frac{64}{27\,N^2} - \frac{28}{27\,N} - \frac{2}{27}\right)\zeta(4) + \left(\frac{320}{9\,N^5} - \frac{400}{27\,N^4} - \frac{440}{81\,N^3} - \frac{80}{9\,N^2} + \frac{280}{81\,N} + \frac{40}{81}\right)\zeta(5)\right] +
\end{aligned}$$

$$
\begin{aligned}
&\epsilon^5 \Big[-\tfrac{7069870976}{1594323\,N^9}+\tfrac{2937809504}{177147\,N^8}-\tfrac{4398801284}{177147\,N^7}+\tfrac{9923276525}{531441\,N^6}-\tfrac{5033294725}{708588\,N^5} \\
&\quad +\tfrac{3132906331}{2834352\,N^4}+\tfrac{256333871}{17006112\,N^3}-\tfrac{264392957}{22674816\,N^2}+\tfrac{4069429}{45349632\,N}-\tfrac{321451}{408146688} \\
&\quad +\Big(-\tfrac{17618048}{6561\,N^8}+\tfrac{46032704}{6561\,N^7}-\tfrac{40407016}{6561\,N^6}+\tfrac{10668718}{6561\,N^5}+\tfrac{4840987}{13122\,N^4}-\tfrac{1176529}{6561\,N^3}+\tfrac{64261}{26244\,N^2}+\tfrac{10361}{104976\,N}-\tfrac{11221}{104976}\Big)\ \zeta(3) \\
&\quad +\Big(\tfrac{75472}{729\,N^6}-\tfrac{125462}{729\,N^5}+\tfrac{8347}{162\,N^4}+\tfrac{7747}{243\,N^3}-\tfrac{72941}{5832\,N^2}-\tfrac{323}{972\,N}-\tfrac{443}{5832}\Big)\zeta(4) \\
&\quad +\Big(-\tfrac{67840}{81\,N^7}+\tfrac{845200}{729\,N^6}-\tfrac{202864}{729\,N^5}-\tfrac{10202}{729\,N^4}-\tfrac{83693}{729\,N^3}+\tfrac{29770}{729\,N^2}+\tfrac{1628}{729\,N}+\tfrac{373}{729}\Big)\zeta(5) \\
&\quad +\Big(\tfrac{3100}{81\,N^5}-\tfrac{1525}{81\,N^4}-\tfrac{425}{162\,N^3}-\tfrac{1025}{81\,N^2}+\tfrac{275}{54\,N}+\tfrac{25}{54}\Big)\zeta(6)+\Big(\tfrac{4312}{27\,N^6}-\tfrac{14896}{27\,N^5}+\tfrac{3626}{9\,N^4}-\tfrac{2254}{27\,N^3}+\tfrac{980}{27\,N^2}-\tfrac{98}{9\,N}-\tfrac{49}{27}\Big)\zeta(7) \\
&\quad +\Big(-\tfrac{12544}{81\,N^7}+\tfrac{13888}{81\,N^6}+\tfrac{712}{27\,N^5}-\tfrac{4354}{81\,N^4}-\tfrac{365}{81\,N^3}+\tfrac{10}{3\,N^2}+\tfrac{85}{81\,N}+\tfrac{11}{81}\Big)\zeta^2(3)\Big] + O[\epsilon^6]\ . \qquad (16)
\end{aligned}
$$

The stability of the fixed points is determined by the eigenvalues $\omega_1$ and $\omega_2$ of the matrix

$$
M_{ij} = \frac{\partial \beta_i(g_1, g_2)}{\partial g_j}\Big|_{g_1^\star, g_2^\star} . \qquad (17)
$$

If the real parts of both eigenvalues are positive, the corresponding fixed point is infrared stable. The Gaussian fixed point is doubly unstable. At the Ising fixed point, one eigenvalue $\omega_1$ is negativ. The Heisenberg and the cubic fixed point interchange stability for $N = N_c$, the former being stable for $N < N_c$ where $g_2^C < 0$. The stability wedges of the critical theory are visible in Fig. 1. They differ from the bare stability wedge. Outside these wedges, the transition is of first order [23].

To find the crucial number $N_c$ determining which fixed point governs the critical behavior in $D = 3$ dimensions, we study the eigenvalue $\omega_2^C$ of the stability matrix as a function of $N$ (the other eigenvalue $\omega_1^C$ remains positive and can be ignored), whose $\epsilon$-expansion reads

$$
\begin{aligned}
\omega_2^C &= \epsilon\,\frac{N-4}{3\,N} + (N-1)\left[\epsilon^2\,\frac{(-848+660\,N+72\,N^2-19\,N^3)}{81\,N^3\,(2+N)}\right. \qquad (18) \\
&\left. + \epsilon^3 \frac{\Sigma_{i=0}^{7}\, C_i^3\, N^i}{8748\,N^5\,(2+N)^3} + \epsilon^4 \frac{\Sigma_{i=0}^{11}\, C_i^4\, N^i}{944784\,N^7\,(2+N)^5} + \epsilon^5 \frac{\Sigma_{i=0}^{15}\, C_i^5\, N^i}{102036672\,N^9\,(2+N)^7}\right] + O[\epsilon^6]\ .
\end{aligned}
$$

The coefficients $C_i^j$ are listed in Table 1. From the vanishing of $\omega_2^C$ we extract the $\epsilon$-expansion of $N_c$

$$
\begin{aligned}
N_c &= 4 - 2\,\epsilon + \epsilon^2\left(-\frac{5}{12}+\frac{5\,\zeta(3)}{2}\right) + \epsilon^3\left(-\frac{1}{72}+\frac{5\,\zeta(3)}{8}+\frac{15\,\zeta(4)}{8}-\frac{25\,\zeta(5)}{3}\right) + \qquad (19) \\
&\epsilon^4\left(-\frac{1}{384}+\frac{93\,\zeta(3)}{128}-\frac{229\,\zeta^2(3)}{144}+\frac{15\,\zeta(4)}{32}-\frac{3155\,\zeta(5)}{1728}-\frac{125\,\zeta(6)}{12}+\frac{11515\,\zeta(7)}{384}\right) + O[\epsilon^5]\ .
\end{aligned}
$$

The same expansion is found from the condition $g_2^C = 0$. Unfortunately, this expansion is badly divergent making it difficult to calculate the value of $N_c$ at $\epsilon = 1$. With the help of Padé approximations we obtain the values

| | | | | | | | |
|---|---|---|---|---|---|---|---|
| Padé [1,1] : | $N_c$ | = | 3.128 | Padé [2,2] : | $N_c$ | = | 2.958 |
| Padé [2,1] : | $N_c$ | = | 2.792 | Padé [1,2] : | $N_c$ | = | 2.893 |
| Padé [3,1] : | $N_c$ | = | 3.068 | Padé [1,3] : | $N_c$ | = | 2.972 |

The approximation Padé [1,3] is unreliable since the Padé denominator has a pole at positive $\epsilon$, where the exact result should be regular. The highest symmetric approximation is usually the most accurate one, from which we deduce the estimate:

$$N_c \approx 2.958. \tag{20}$$

Before our work, the $\epsilon$-expansion for $N_c$ was known only up to the order $\epsilon^2$ [7] so that only the Padé [1,1]-approximation was available [7] which yielded $N_c \approx 3.128$. An alternative method, the scaling field method of [24] in which the RG-equations are transformed into an infinite set of differential equations, was used in Ref. [25] to estimate $N_c$ directly in $D = 3$ dimensions yielding $N_c = 3.4$. Thus the previous results yielded $N_c$-values larger than 3 implying that the critical behavior of magnetic systems with cubic symmetry is governed by the Heisenberg fixed point. The symmetric Padé approximation to our expansion suggests that $N_c$ lies below three so that the cubic fixed point is the relevant one.

By inserting the expansions (15) and (16) into (12) and (13), we find the critical exponents $\eta_C$ and $\nu_C$ of the cubic fixed point via the defining relations [4]

$$\eta = 2\gamma_2(g_1^\star, g_2^\star) \ , \quad 1/\nu = 2(1 - \gamma_m(g_1^\star, g_2^\star)) \ . \tag{21}$$

The exponents are

$$1/\nu^C = 2+(N-1)\Big\{-\epsilon\,\tfrac{2}{3\,N}+\epsilon^2\left[\tfrac{1}{162\,N^3}\left(424-326\,N+19\,N^2\right)\right]+$$
$$\epsilon^3\left[\tfrac{1}{17496\,N^5}\left(-359552+573728N-264936N^2+28358N^3+937N^4+\tfrac{4\,(N+2)}{27\,N^4}\left(-14+11\,N-N^2\right)\zeta(3)\right)\right]+$$
$$\epsilon^4\left[\tfrac{1}{1889568\,N^7}\left(381125120-923268480N+798088608N^2-284926360N^3+32693424N^4+768780N^5+24857N^6\right)\right.$$
$$+\tfrac{1}{1458\,N^6}\left(118720-152032N+29816N^2+17936N^3-4124N^4-119N^5\right)\zeta(3)$$
$$\left.+\tfrac{(2+N)}{9\,N^4}\left(-14+11\,N-N^2\right)\zeta(4)+\tfrac{40}{81\,N^5}\left(36-2\,N-4\,N^2-8\,N^3+N^4\right)\zeta(5)\right]+$$
$$\epsilon^5\left[\tfrac{1}{204073344\,N^9}\left(-452471742464+1470211004416N-1869697955840N^2+1160186503168N^3\right.\right.$$
$$\left.-350446218272N^4+41122747144N^5+144762448N^6-68383472N^7+64327N^8\right)$$
$$+\tfrac{1}{52488\,N^8}\left(-70472192+153589248N-106996288N^2+18129888N^3+6458072N^4-1726592N^5-8716N^6-3335N^7\right)\zeta(3)$$
$$+\tfrac{1}{1944\,N^6}\left(118720-152032N+29816N^2+17936N^3-4124N^4-119N^5\right)\zeta(4)$$

$$+\frac{1}{2187\,N^7}\left(-915840+897560N-53320\,N^2-2676\,N^3-86879\,N^4+20128\,N^5+528\,N^6\right)\zeta(5)$$

$$+\frac{50}{81\,N^5}\left(36-2\,N-4\,N^2-8\,N^3+N^4\right)\zeta(6)+\frac{49}{27\,N^6}\left(44-132\,N+65\,N^2-9\,N^3+8\,N^4-N^5\right)\zeta(7)$$

$$+\frac{8}{81\,N^7}\left(-784+588\,N+232\,N^2-163\,N^3-23\,N^4+5\,N^5+N^6\right)\zeta^2(3)\Big]\Bigg\}+O[\epsilon^6]\;. \qquad (22)$$

$$\eta^C = (N-1)\Bigg\{\epsilon^2\left[\frac{(2+N)}{54\,N^2}\right]+\epsilon^3\left[\frac{-1696+1728\,N-222\,N^2+109\,N^3}{5832\,N^4}\right]+$$

$$\epsilon^4\Big[\frac{1}{629856\,N^6}\left(1797760-3566912N+2292328N^2-507952\,N^3+28832\,N^4+7217\,N^5\right)$$

$$+\frac{4}{243\,N^5}\left(28-6\,N-16\,N^2+4\,N^3-N^4\right)\zeta(3)\Big]+$$

$$\epsilon^5\Big[\frac{1}{68024448\,N^8}\big(-2134300672+6125897728N-6643967232N^2+3326175872N^3$$

$$-731940728N^4+46139232N^5+1948700N^6+321511N^7\big)$$

$$+\frac{1}{17496\,N^7}\left(-189952+266624N-31584N^2-80376N^3+29704N^4-2196\,N^5-329\,N^6\right)\zeta(3)$$

$$+\frac{1}{81\,N^5}\left(28-6\,N-16\,N^2+4\,N^3-N^4\right)\zeta(4)+\frac{40}{729\,N^6}\left(-36+N^2+18\,N^3-5\,N^4+N^5\right)\zeta(5)\Big]\Bigg\}+O[\epsilon^6]\;. \qquad (23)$$

As anticipated, the exponents for the cubic fixed point, are degenerate with those of the Gaussian fixed point, taking free-field values. At $N=2$, the exponents take Ising values.

For a reliable extraction of the critical exponents from our expansions it will be necessary to find the large-order behavior of the perturbation expansions in $g_1$ and $g_2$. A combination of Borel and Padé resummation methods are expected to give the best results. Such methods should confirm the result $N_c < 3$ and render the precise differences between the critical exponents of the Heisenberg and the cubic fixed point. From the above rough Padé resummations, we expect the differences to be less than a percent. This will make it difficult distinguishing the two fixed points experimentally.

| | | | |
|---|---|---|---|
| $C_0^3$ | $2876416$ | $C_2^4$ | $22268920832-5633556480\zeta(3)-1074954240\zeta(5)$ |
| $C_1^3$ | $-1740544+580608\zeta(3)$ | $C_3^4$ | $-5244000000+5418233856\zeta(3)+188116992\zeta(4)-2120048640\zeta(5)$ |
| $C_2^3$ | $-3188544+829440\zeta(3)$ | $C_4^4$ | $-21313343616+8161855488\zeta(3)+456855552\zeta(4)-2508226560\zeta(5)$ |
| $C_3^3$ | $2340592-62208\zeta(3)$ | $C_5^4$ | $11104506624-370593792\zeta(3)+295612416\zeta(4)-1425807360\zeta(5)$ |
| $C_4^3$ | $54656-404352\zeta(3)$ | $C_6^4$ | $1087015200-2657484288\zeta(3)-83980800\zeta(4)+432967680\zeta(5)$ |
| $C_5^3$ | $-162696-114048\zeta(3)$ | $C_7^4$ | $-1171729344-359023104\zeta(3)-173000448\zeta(4)+901393920\zeta(5)$ |
| $C_6^3$ | $15700+7776\zeta(3)$ | $C_8^4$ | $54071304+211455360\zeta(3)-67184640\zeta(4)+343388160\zeta(5)$ |
| $C_7^3$ | $-937+2592\zeta(3)$ | $C_9^4$ | $26339632+27133056\zeta(3)-5878656\zeta(4)+20062080\zeta(5)$ |
| $C_0^4$ | $-12196003840$ | $C_{10}^4$ | $-3320774-2164320\zeta(3)+1469664\zeta(4)-9797760\zeta(5)$ |
| $C_1^4$ | $5267640320-4923555840\zeta(3)$ | $C_{11}^4$ | $-24857+154224\zeta(3)+209952\zeta(4)-933120\zeta(5)$ |

| | |
|---|---|
| $C_0^5$ | $57916383035392$ |
| $C_1^5$ | $-14984240431104+35071472959488\zeta(3)$ |
| $C_2^5$ | $-143277741441024+39325680009216\zeta(3)+2022633897984\zeta^2(3)+10938734346240\zeta(5)$ |
| $C_3^5$ | $1211449212928-58923130945536\zeta(3)+5995664769024\zeta^2(3)$ $-1595232092160\zeta(4)+22263974461440\zeta(5)-2085841207296\zeta(7)$ |
| $C_4^5$ | $162125071257600-92450858139648\zeta(3)+5046265184256\zeta^2(3)-3420504391680\zeta(4)$ $+9499251179520\zeta(5)-580475289600\zeta(6)+3887249522688\zeta(7)$ |
| $C_5^5$ | $38426107425792+21006186430464\zeta(3)-2702434959360\zeta^2(3)-468572553216\zeta(4)$ $-11373660831744\zeta(5)-1725301555200\zeta(6)+14600888451072\zeta(7)$ |
| $C_6^5$ | $-162464030196224+69899709652992\zeta(3)-6924425232384\zeta^2(3)+3943630872576\zeta(4)$ $-19961756909568\zeta(5)-2644387430400\zeta(6)+13700184293376\zeta(7)$ |
| $C_7^5$ | $53925927185664+4311047245824\zeta(3)-3110702579712\zeta^2(3)+2963245731840\zeta(4)$ $-11389462659072\zeta(5)-2410584883200\zeta(6)+5736063320064\zeta(7)$ |
| $C_8^5$ | $14422298978304-19789519380480\zeta(3)+993580204032\zeta^2(3)-319987003392\zeta(4)$ $+1995616714752\zeta(5)-874744012800\zeta(6)+592568524800\zeta(7)$ |
| $C_9^5$ | $-7438725755776-2538882164736\zeta(3)+1199003959296\zeta^2(3)-1007366492160\zeta(4)$ $+4815696457728\zeta(5)+528071270400\zeta(6)-1724374407168\zeta(7)$ |
| $C_{10}^5$ | $-147321611712+2076145468416\zeta(3)+251458670592\zeta^2(3)-263068176384\zeta(4)$ $+1308067024896\zeta(5)+730632960000\zeta(6)-2005844456448\zeta(7)$ |
| $C_{11}^5$ | $258021138336+211760262144\zeta(3)-41520107520\zeta^2(3)+48221775360\zeta(4)$ $-269590685184\zeta(5)+317951308800\zeta(6)-960701720832\zeta(7)$ |
| $C_{12}^5$ | $-8428414672-68664506112\zeta(3)-17716589568\zeta^2(3)+25217754624\zeta(4)$ $-141247120896\zeta(5)+51900134400\zeta(6)-170733806208\zeta(7)$ |
| $C_{13}^5$ | $-2039889852-1386414144\zeta(3)-851565312\zeta^2(3)+1546506432\zeta(4)$ $-7514135424\zeta(5)-3086294400\zeta(6)+12221725824\zeta(7)$ |
| $C_{14}^5$ | $-138029874+936160416\zeta(3)-25194240\zeta^2(3)-125341344\zeta(4)+1017287424\zeta(5)-1826582400\zeta(6)+6666395904\zeta(7)$ |
| $C_{15}^5$ | $-64327+12966480\zeta(3)-20155392\zeta^2(3)+12492144\zeta(4)-49268736\zeta(5)-125971200\zeta(6)+370355328\zeta(7)$ |

Table 1: The constants appearing in Eq. (19) for $\omega_2^C$.